\documentclass[twocolumn,floatfix]{aastex631}
\usepackage[ngerman,english]{babel}
\newif\ifAMStwofonts
\AMStwofontstrue

\def\bz{{b(z)}}
\def\nz{{N(z)}}
\def\cl{{C_\ell}}

\def\clth{{C^{\rm th}_\ell}}
\def\clobs{{C^{\rm obs}_\ell}}

\def\al{{a_{\ell m}}}

\def\gsim{~\rlap{$>$}{\lower 1.0ex\hbox{$\sim$}}}

\def\simpropto{\lower.2ex\hbox{$\; \buildrel \propto \over \sim \;$}}
\def\ltsim{\lower.5ex\hbox{$\; \buildrel < \over \sim \;$}}
\def\gtsim{\lower.5ex\hbox{$\; \buildrel > \over \sim \;$}}
\def\ltsim{\lower.5ex\hbox{$\; \buildrel < \over \sim \;$}}
\def\gtsim{\lower.5ex\hbox{$\; \buildrel > \over \sim \;$}}

\def\H{{\cal H}}

\def\s{\,{\rm s}}
\def\K{\,{\rm K}}

\def\sun{\Theta_{_{\rm un}}}

\def\dd{\,{\rm d}}

\def\vb{{v_{_{\rm b}}}}

\def\dd{{\rm d}}

\def\ln{{\rm ln}}

\def\pmb#1{\setbox0=\hbox{#1}%
\kern-.025em\copy0\kern-\wd0
\kern.05em\copy0\kern-\wd0
\kern-.025em\raise.0433em\box0}

\def\vr{\pmb{r}}

\def\hvn{\hat {\vr}}

\def\vk{\pmb{k}}

\def\simlt{\lower.5ex\hbox{$\; \buildrel < \over \sim \;$}}
\def\simgt{\lower.5ex\hbox{$\; \buildrel > \over \sim \;$}}

\newcommand{\beq}{\begin{equation}}

\newcommand{\eeq}{\end{equation}}
\def\beqa{\begin{eqnarray}}
\def\eeqa{\end{eqnarray}}
\def\fixit#1{}

\def\dd{{\rm d}}

\def\apjs{ApJ. S}
\def\cN{{\cal N}}

\def\apjl{ApJL}

\def\aap{A\&A}

\def\apjl{ApJL}

\usepackage[normalem]{ulem}
\usepackage{soul}
\usepackage{braket}
\usepackage{physics}

\graphicspath{{./}{figures/}}
\begin{document}

\title{The clustering properties of AGNs/quasars in CatWISE2020 catalog }
\author[0000-0001-7888-4270]{Prabhakar Tiwari}
\affiliation{National Astronomical Observatories,
Chinese Academy of Science, Beijing, 100101, P.R.China}
\author[0000-0003-4726-6714]{Gong-Bo Zhao}
\affiliation{National Astronomical Observatories,
Chinese Academy of Science, Beijing, 100101, P.R.China}
\author[0000-0002-8272-4779]{Adi Nusser}
\affiliation{Department of Physics and the Asher Space Research Institute, Technion, Haifa 3200003, Israel}

\begin{abstract}
We study the clustering properties of 1,307,530 AGNs/quasars in the CatWISE2020 catalog prepared using the Wide-field Infrared Survey Explorer (WISE) and Near-Earth Object Wide-field Infrared Survey Explorer (NEOWISE) survey data. For angular moments $\ell \gtrapprox 10$ ($\lessapprox 18^\circ$) down to non-linear scales, the results are in agreement with the standard $\Lambda$CDM cosmology,  with a galaxy bias roughly matching that of the NRAO VLA Sky Survey (NVSS) AGNs.  We further explore the redshift dependence of the fraction of infrared bright AGNs on stellar mass, $f_{\rm IB} \sim M_*^{\alpha_0 + \alpha_1 z}$, and find $\alpha_1=1.27^{+0.25}_{-0.30}$, ruling out a non-evolution hypothesis at $\approx 4.6\sigma$ confidence level. The results are consistent with the measurements obtained with NVSS AGNs, though considerably more precise thanks to the significantly higher number density of objects in CatWISE2020. The excess dipole and high clustering signal above angular scale $\approx 18^\circ$ remain anomalous. 
\end{abstract}
\keywords{Cosmology (343); Large-scale structure of the universe (902);  Observational cosmology (1146); Infrared astronomy (786); Quasars (1319); Active galactic nuclei (16)}
\section{Introduction}
\label{sc:intro}

Recently, \cite{Secrest:2020CPQ} derived an all-sky active galactic nuclei (AGN)/quasar sample from the CatWISE2020 catalog (AGN-CatWISE2020 hereafter) and studied the large-scale anisotropy of the universe by measuring the dipole signal. They find a dipole amplitude that conflicts with the prediction of the $\Lambda$CDM at the $4.9\sigma$ level,
as had been also indicated, albeit with less significance, in earlier studies of the dipole in AGN catalogs prepared using radio telescopes \citep{Singal:2011,Rubart:2013,Tiwari:2014ni,Tiwari:2015np, Tiwari:2016adi, Siewert:2020CRD}.

The AGN-CatWISE2020 is a satellite-based catalog and presumably not affected by systematics and observational biases such as declination-dependent sensitivity or atmospheric effects present in catalogs prepared using ground-based radio surveys. However, the  AGN-CatWISE2020, prepared using deep photometric measurements in the infrared band at $3.4$ and $4.6$ $\mu$m  from the cryogenic, post-cryogenic, and reactivation phases of the WISE mission has its own observational systematics, such as the directional bias from uncorrected Galactic reddening, poor-quality photometry near clumpy and resolved nebulae, image artifacts near bright stars, uneven source density due to scanning overlap etc. Even so, the AGN-CatWISE2020 remains an independent AGN catalog with a completely different observational setup and instrumentation in contrast to canonical radio AGN catalogs prepared using ground-based radio telescope surveys, and thus provides the opportunity to independently study AGN spatial distribution properties and their connection to background matter density and their host halos. 

Our present picture of structure formation is based on the standard $\Lambda$ cold dark matter ($\Lambda$CDM); the model describes  the large-scale structure formation and its evolution successfully. The matter density of the Universe is dominated by cold dark matter, 
and according to the  theory of gravitational instability, tiny fluctuations in the initial mass density field evolve  and result in a population of virialized dark matter halos of different masses \citep{Press:1974}. The formation of galaxies occurs inside these dark matter halos, the host halo mass and its evolution largely decide the residing galaxy properties \citep{White:1978,Frenk:1988,White:1991}. The AGNs span the high mass end  of the halo mass distribution and thus with AGN clustering analysis, we explore the clustering properties of relatively high mass halos. The radio emission from a galaxy and its stellar mass is known to have a close connection \citep{Best:2005,Donoso:2009, Adi:2015nb,Sabater:2019}. We assume a similar connection for infared selected AGNs and  further explore this connection in this work. 

We follow the standard correlation function and power spectrum analysis to explore the spatial distribution of AGNs from the WISE mission and their connection to the background dark matter distribution, i.e., the galaxy bias for these AGNs. Throughout the paper, we adopt the  $\Lambda$CDM model and use Planck $2018$ cosmological parameters as the fiducial cosmology. In particular, we set base cosmology parameters from \cite{Planck_results:2018}, i.e. TT,TE,EE+lowE+lensing $\omega_b h^2 =0.02237$,    $\omega_c h^2  =0.1200$,  $100\theta_{\rm MC}=1.04092$, $\tau=0.0544$,   $\ln (10^{10} A_s)=3.044$ and $n_s=0.9649$.

We discuss the data and AGN selection details in Section \ref{sc:data}. In Section \ref{sc:theory}, we  briefly describe the theoretical formulation for angular power spectrum and projected two-point correlation function.  Mocks data and and covariance matrix generation are detailed  in Section \ref{sc:mocks}.  Section \ref{sc:results} presents the results. We conclude with discussion in Section \ref{sc:conclusion}. 

\section{CatWISE2020 catalog and AGN selection}
\label{sc:data}
\begin{figure}
    \centering
    \includegraphics[scale=0.4]{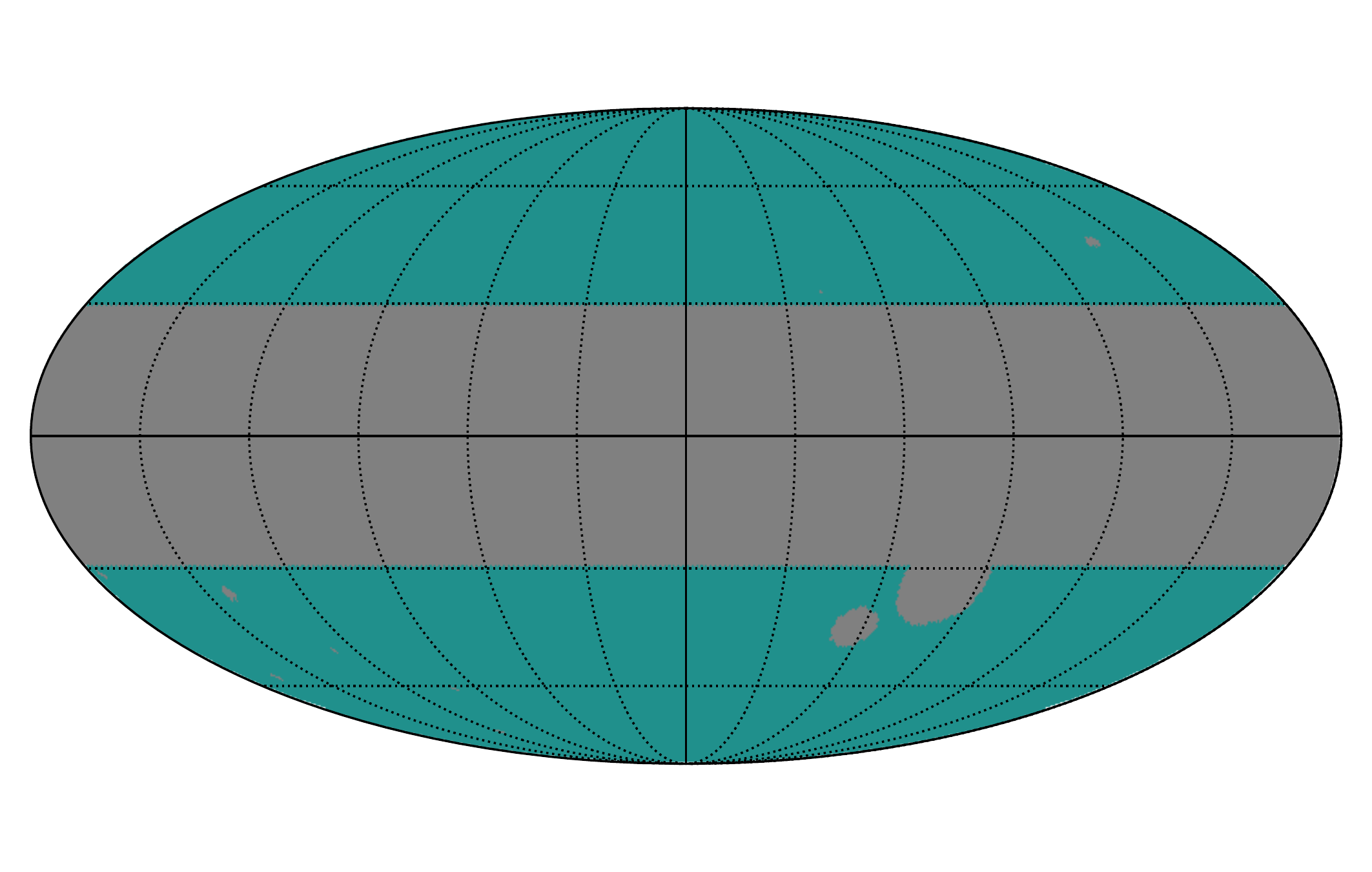}
    \includegraphics[scale=0.4]{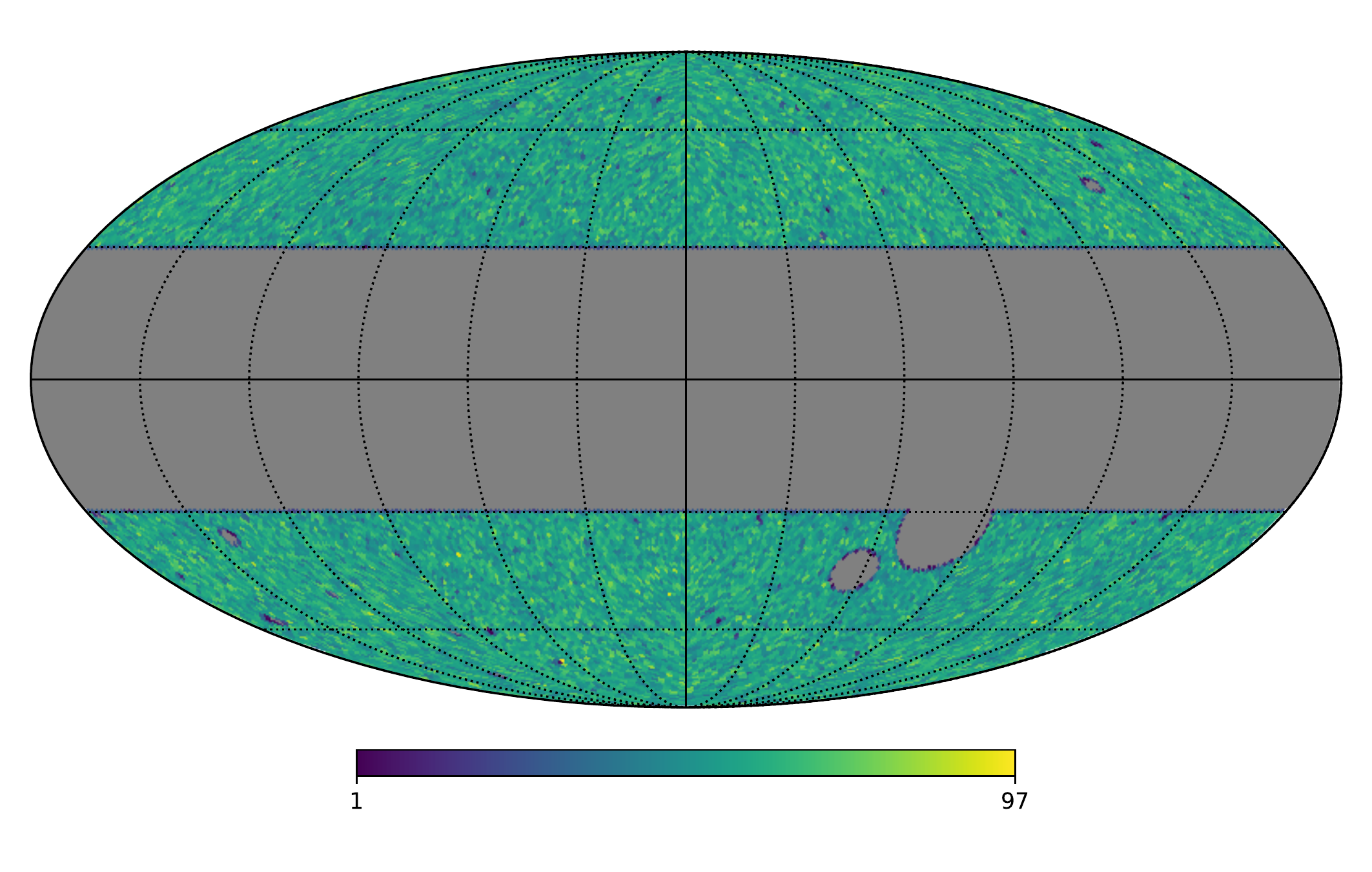}
    
    \caption{AGN-CatWISE2020 mask (top) and catalog by \cite{Secrest:2020CPQ} in galactic coordinates. There are 1,355,352 sources in the catalog and the data cover 49.65\% of the sky. }
    \label{fig:AGNcat0}
\end{figure}

\begin{figure}
    \centering
    \includegraphics[scale=0.4]{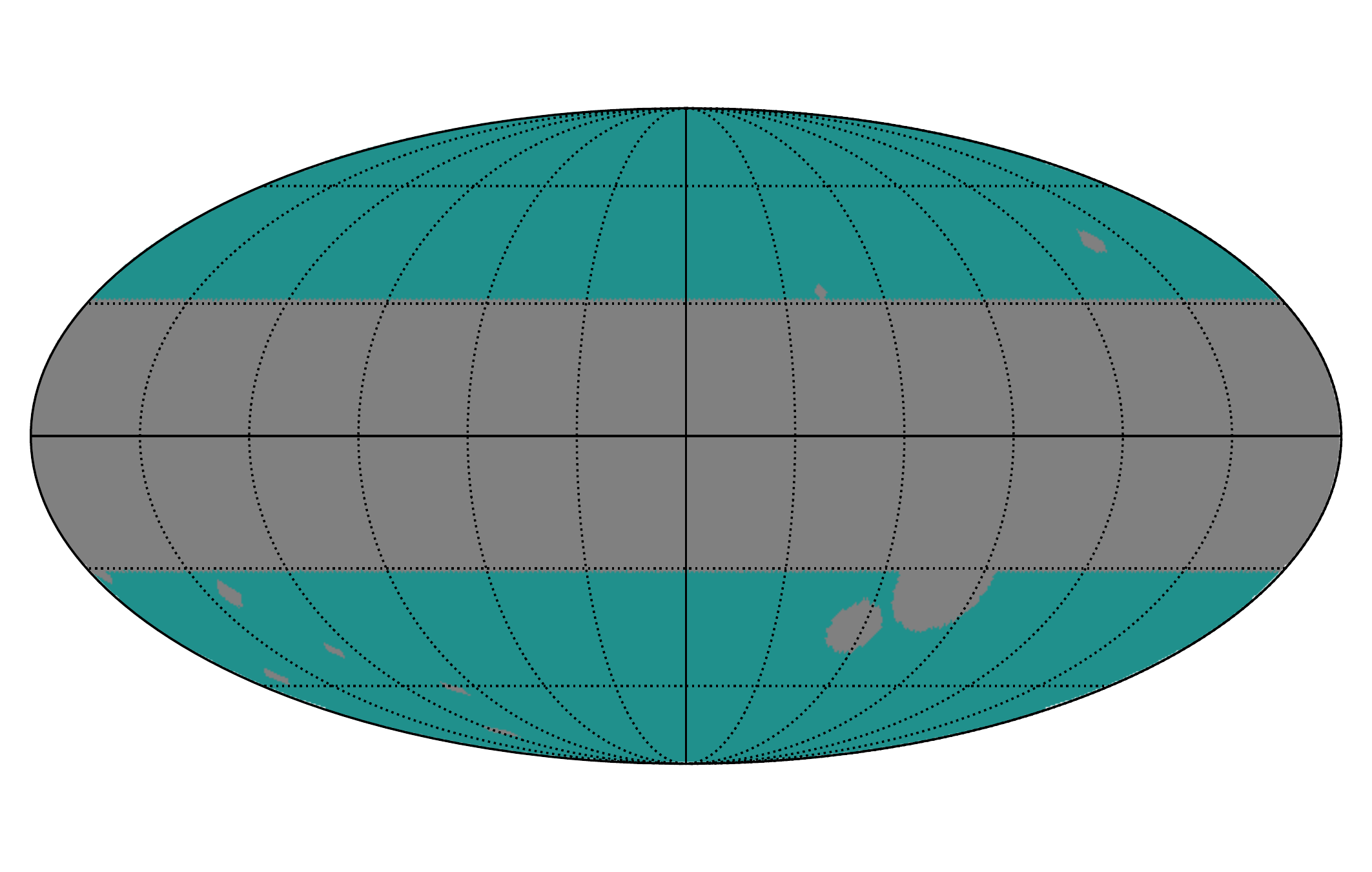}
    \includegraphics[scale=0.4]{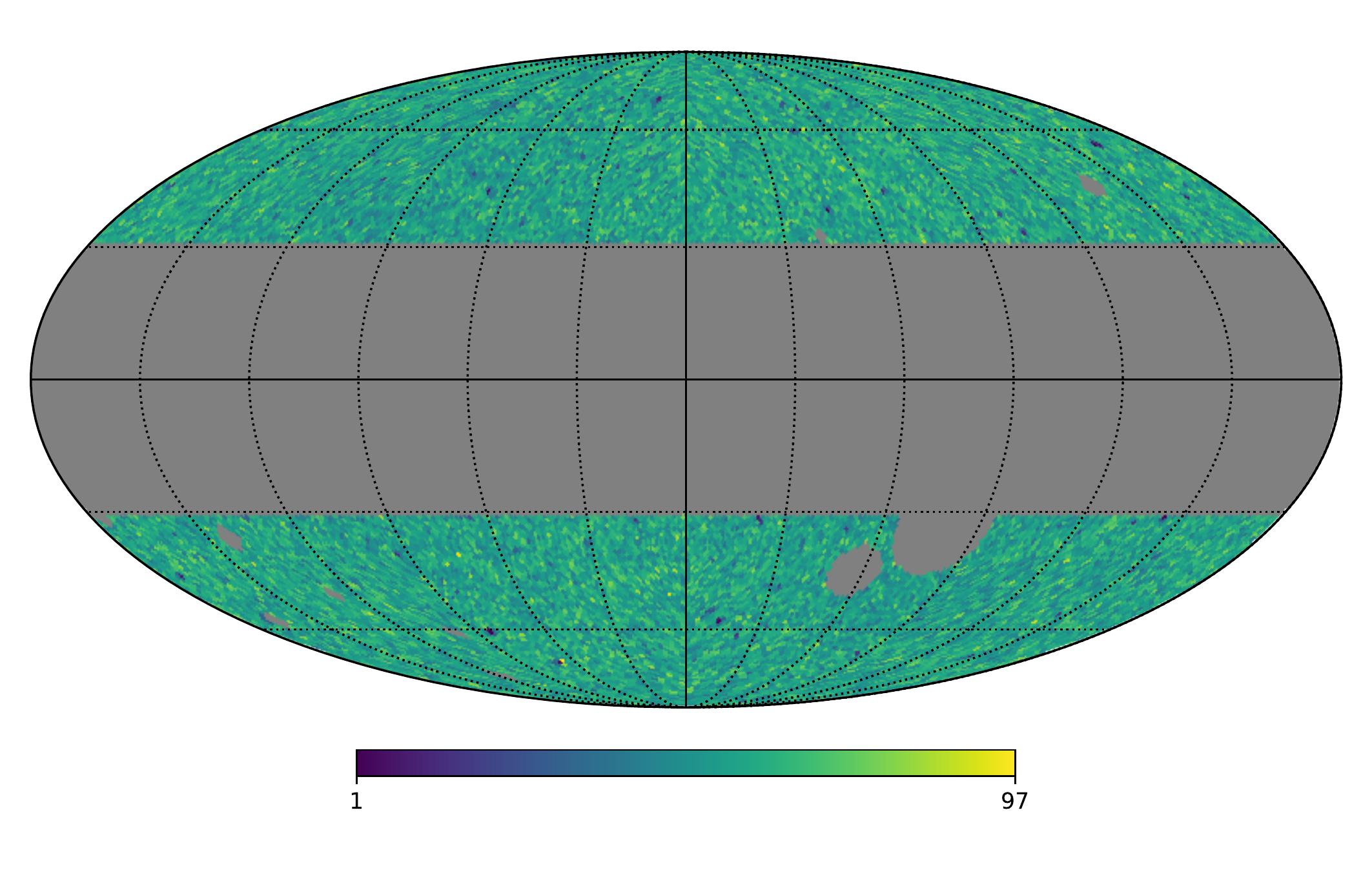}
    \caption{Alternate mask and AGN catalog after masking in galactic coordinates. We are left with 1,307,530 sources in the catalog and the data covers 47.35\% of the sky.}
    \label{fig:AGNcat}
\end{figure}

\begin{figure}
    \centering
    \includegraphics[scale=0.32]{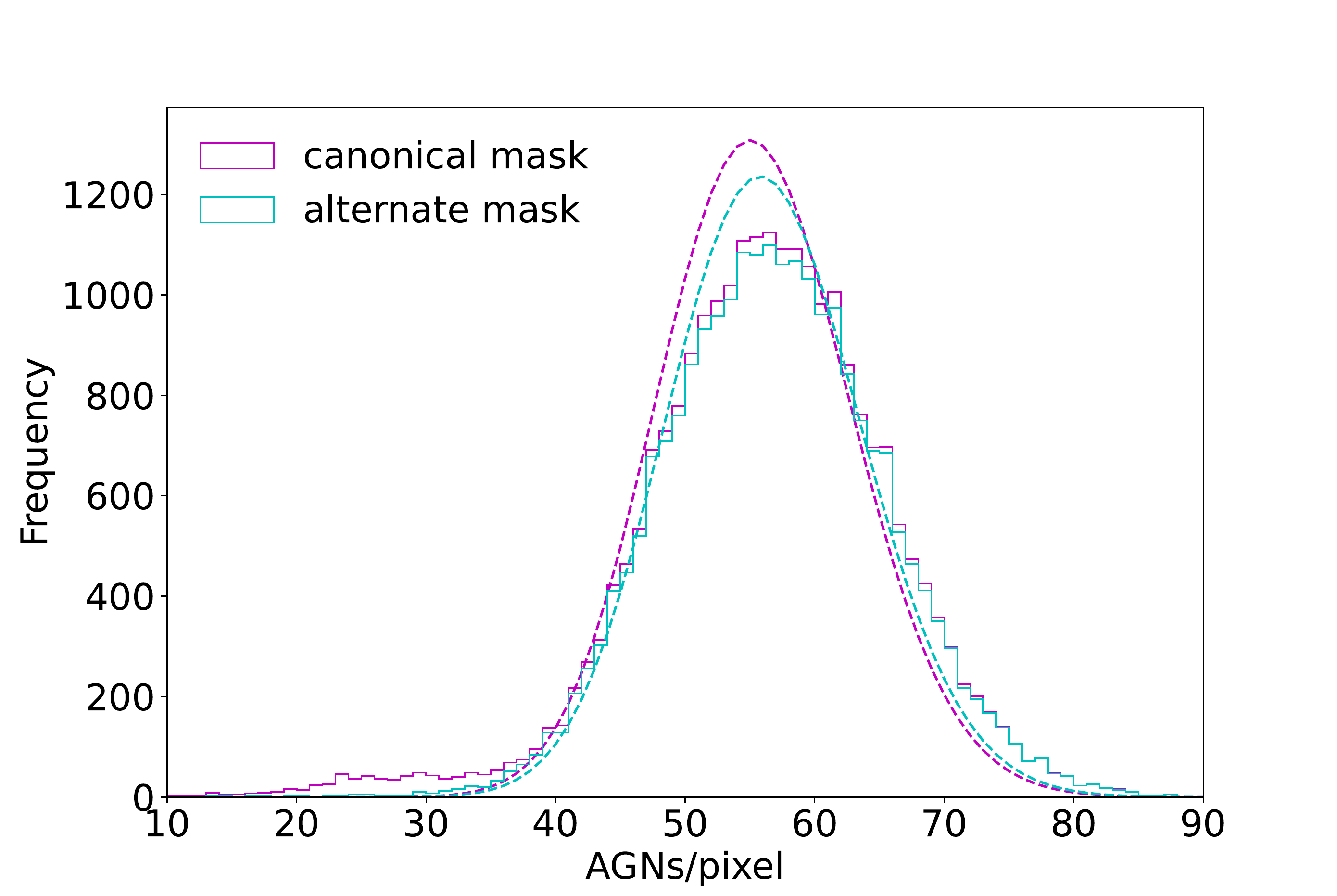}
    \caption{The source count in {\tt HEALPix} pixels with {\tt NSIDE=64}. The canonical mask shows a tail towards zero ends. The counts in pixels with our alternate mask are a better fit with the Poisson distribution (dashed line curves).}
    \label{fig:pixcount}
\end{figure}

The CatWISE2020 \citep{CatWise2020:2021} catalog covers the entire sky and contains 1,890,715,640 sources from the Wide-field Infrared Survey Explorer (WISE, \citealt{Wright:2010}) and the Near-Earth Object Wide-field Infrared Survey Explorer (NEOWISE) deep photometric surveys at 3.4 and 4.6 $\mu$m  (W1 and W2). The catalog is 90\% complete at W1=17.7 mag and W2=17.5 mag. Relying on distinguishing the power-law spectrum of AGNs from the blackbody spectrum of galaxies and stars, the mid-infrared WISE bands W1 and W2 can reliably identify AGNs. A simple color criterion of $W1-W2 \ge 0.8$ identifies both unobscured (type 1) and obscured (type 2) AGNs \citep{Stern:2012} which presumably follow their characteristic power-law ($S_\nu \propto \nu^{-\alpha}$). \cite{Secrest:2020CPQ} apply this color criterion, i.e.,  $W1-W2 \ge 0.8$ on CatWISE2020 catalog and find  $141,698,603$ sources. They mask the nebulae in our and nearby galaxies and the bright stars (in total $291$ sky regions) to remove the poor-quality photometry regions. Also, they impose a magnitude selection cut of $9<W1<16.4$ to correct for uneven source density due to overlap in the WISE scanning pattern. Further they mask Galactic latitudes of $|b| < 30^\circ$ and impose some additional  masks and cuts (for details see \citealt{Secrest:2020CPQ}) and produce a source catalog of $1,355,352$ sources. The source catalog thus obtained and the mask are shown, using {\tt HEALPix}\footnote{\href{https://healpix.sourceforge.io}{https://healpix.sourceforge.io}} \citep{Gorski:2005} Mollweide projection with {\tt NSIDE=64}, in Figure \ref{fig:AGNcat0}. The catalog covers almost 50\% of the sky. We note that the projected pixel count in cells shows a tail toward zero ends (Figure \ref{fig:pixcount}). We find this to be because of low source count at mask edges, an artifact of projection in {\tt HEALPix} pixels of finite sizes. We thus additionally mask all the neighbouring pixels of masked pixels and produce an alternate mask for AGN-CatWISE2020. With this alternate mask we are left with 1,307,530 sources and around 47.25\% of sky coverage. The mask and the projected source number density is shown in Figure \ref{fig:AGNcat}. For our all analysis in this work we use this alternate mask and 1,307,530 AGNs left after applying this mask. The source count in pixels with canonical and our alternate mask is shown in Figure \ref{fig:pixcount}. The CatWISE catalog also shows a mild inverse linear trend in source number density versus absolute ecliptic latitude with  a slope of $-0.051$ and a zero-latitude intercept of $68.89$ deg$^{-2}$\citep{Secrest:2020CPQ}. We correct the number density to account for this trend. All results throughout the paper are after applying the ecliptic latitude correction.

\subsection{The redshift distribution}
\label{ssc:nz}
\begin{figure}[ht]
    \centering
    \includegraphics[scale=0.6]{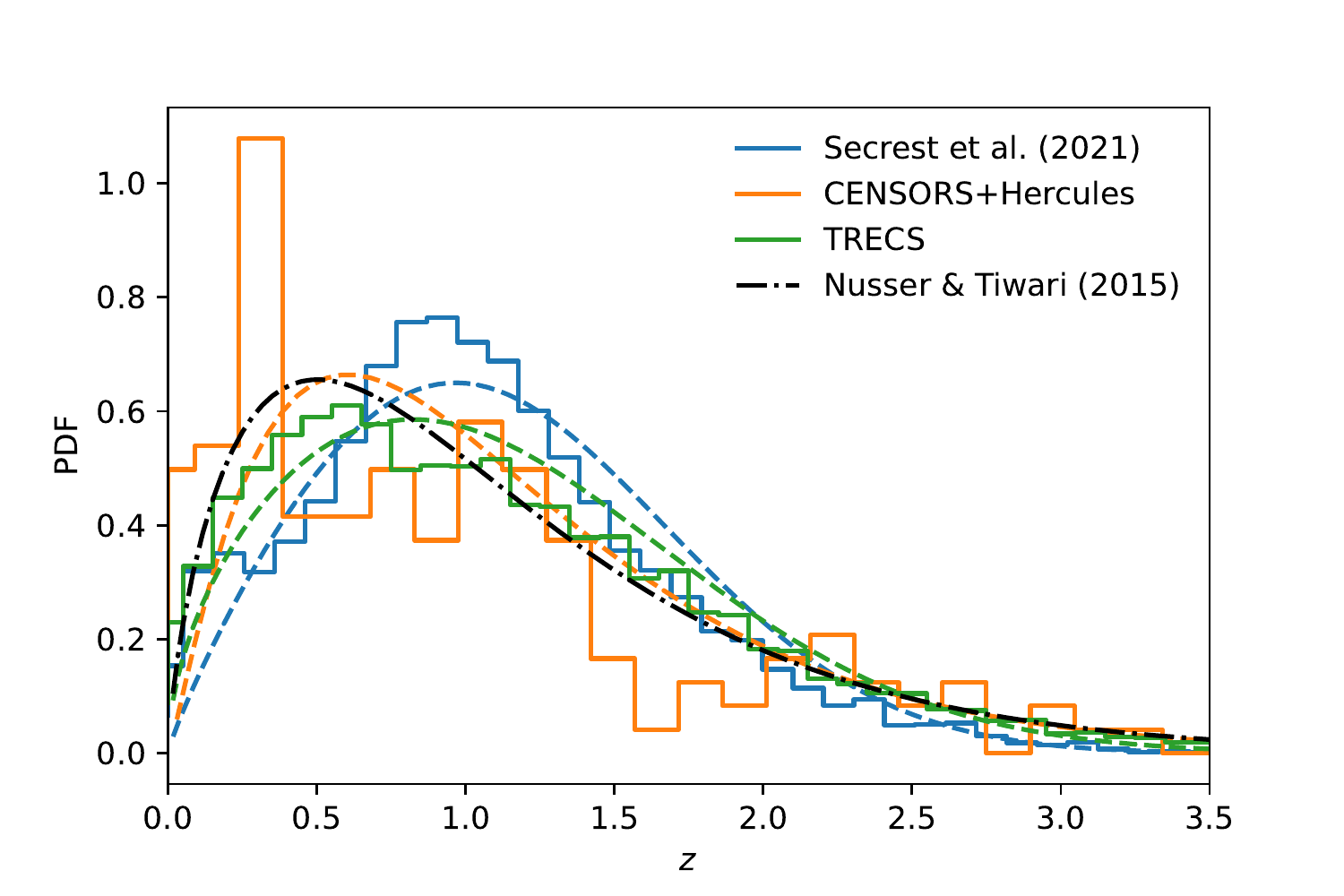}
    \caption{The redshift distribution of AGN-Catwise2020 sample by \cite{Secrest:2020CPQ}. We also show the best fit PDF to the NVSS AGNs with integrated flux grater than 10 mJy at 1.4 GHz and the TRECS for comparison. The best fitted (dashed lines), parameterized $N(z) \propto z^{a_1} \exp\left[ - \left(\frac{z}{a_2} \right)^{a_3} \right]$, are also shown for different $N(z)$ templates.}
\label{fig:zpdf}
\end{figure}

The photometric data from the WISE and NEOWISE do not contain redshift information. To derive an estimate for their  redshift distributions,  \cite{Secrest:2020CPQ} cross match the catalog with SDSS strip 82 sub-region between $-42^\circ <$ R.A. $<45^\circ$, this area lies outside the AGN-CatWISE2020 mask and has been surveyed several times by SDSS and by 
the extended Baryon Oscillation Spectroscopic Survey (eBOSS; \citealt{Dawson:2016}). This region also has an overlap with the  Dark Energy Survey, Data Release 1 (DES1; \citealt{Abbott:2018}) and Gaia Data Release 2 \citep{GaiaDR2:2018}. \cite{Secrest:2020CPQ} report 14,402 AGNs in this region and after correcting  DES1 source positions using Gaia DR2 they find DES1 counterparts for 14,193 sources (99\%). 
They find spectroscopic information for $\sim 61$\% of the AGNs using specObj tables for SDSS DR16\footnote{\href{https://www.sdss.org/dr16/spectro/spectro_access}{https://www.sdss.org/dr16/spectro/spectro\_access}}. The redshift distribution thus obtained is shown in Figure \ref{fig:zpdf}. Using DES1 and WISE photometric bands, i.e., r-W2 value  \cite{Secrest:2020CPQ} argue that the AGNs without SDSS spectral redshifts are faint at visual wavelengths. The matched objects may be used to represent the distribution of redshifts for the full sample. However, as this redshift PDF is prepared using
spectroscopic information of only $\sim 61$\% sources,  considering this as reliably representing full sample remains debatable.

We therefore produce an alternative redshift template for AGN-CatWISE2020 employing the Tiered Radio Extragalactic Continuum Simulation (TRECS; \citealt{Bonaldi:2018}). We run TRECS for sub-mJy flux limits (at 1.4GHz) and produce redshifts for flux limited data. We  find that TRECS AGNs number density matches with AGN-CatWISE2020 if we consider flux density $>1.15$ mJy at 1.4 GHz.
Assuming, AGN-CatWISE2020 is complete above a constant flux density, the TRECS redshifts above flux density $>1.15$ mJy at 1.4 GHz represent AGN-CatWISE2020. We have shown the redshift PDF thus obtained using TRECS in Figure \ref{fig:zpdf}. For comparison, we also show the PDF best fitted to NRAO VLA Sky Survey (NVSS; \citealt{Condon:1998}) AGNs \citep{Adi:2015nb}. This best fit to NVSS by  \cite{Adi:2015nb} is obtained by considering the redshifts from Combined EIS-NVSS Survey of Radio Source (CENSORS) and Hercules \citep{Rigby:2011},  both in total containing  165 sources above 7.5 mJy at 1.4 GHz. We have shown CENSORS+Hercules redshifts (normalized) also in same figure for comparison. We will also use this redshift distribution as an alternate template to fit CatWISE2020 AGNs  considered in this work. We add that TRECS and CENSORS+Hercules represent the radio-selected AGNs and WISE-selected AGNs may have somewhat different $\nz$. We employ these as optional $\nz$. In any case, these redshift distributions only slightly differ from \cite{Secrest:2020CPQ} (see Figure \ref{fig:zpdf}).
\section{Theoretical formulation}
\label{sc:theory}
In the absence of redshift measurements, the angular power spectrum and angular correlation function of galaxies are  standard tools for analyzing 
the clustering properties of galaxies and inferring information on their luminosity, biasing, and background cosmology \citep{Magliocchetti:1999,Blake:2002ac,Overzier:2003,Adi:2015nb,Hale:2018,Tiwari:2021DR1}. We briefly discuss these below. 

\subsection{Angular power spectrum}
\label{ssc:cls}
Given a uniform galaxy survey catalog with a total number of galaxies $\cN$ and with angular area coverage $\mathcal{A}$, the mean number density per steradian is,  $\bar \cN=\cN/\mathcal{A}$. The projected number density in the direction $\hvn$, can have the expression,  $\cN(\hvn) =\bar \cN(1+\Delta (\hvn))$, where $\Delta (\hvn)$ is the projected number density contrast and is theoretically connected to the underlying matter density contrast, $\delta_m(\vr)$. The galaxy density contrast $\delta_g(\vr)$ traces  $\delta_m(\vr)$ through a bias factor, the galaxy biasing $b(z)$, 
\beq
\delta_g(\vr) =\delta_m(\vr) b(z),
\eeq
Note $\vr$ stands for comoving distance $r$ corresponding to redshift $z$ in direction $\hat{r}$. $\Delta (\hvn)$ can be obtained by integrating over the radial distribution of galaxies. 
\begin{eqnarray}
\label{eq:delta_th}
\Delta (\hvn) &=& \int _{0}^{\infty} \delta_g(\vr)N(z) dz  \nonumber\\
              &=&  \int _{0}^{\infty} \delta_m(\vr) b(z) N(z) dz ,
\end{eqnarray}
where $N(z) \dd z$, the radial distribution function, is the probability of observing a galaxy between redshift $z$ and $(z+ \dd z)$. In addition to this leading term,  $\Delta (\hvn)$ also have tiny additional contributions from lensing, redshift distortions etc. \citep{Chen:2015}. These effects are within a few percent on the largest scales \citep{Dolfi:2019}. To estimate the galaxy clustering at different angular scales, we expand $\Delta (\hvn)$ in terms of spherical harmonics,
\beq
\label{eq:alm}
\Delta (\hvn) = \sum_{\ell m}\al Y_{\ell m}(\hvn).
\eeq
Using the orthonormal property of spherical harmonics, we write $\al$ as, 
\begin{eqnarray}
\label{eq:alm_gal}
\al&=&\int d \Omega \Delta(\hvn)  Y_{\ell m}^*(\hvn)\\
\nonumber &=& \int d\Omega Y_{\ell m}^*(\hvn) \int_{0}^{\infty}  \delta_m(\vr)  b(z) N(z) dz\;.
\end{eqnarray}

Next one can Fourier transform the matter density field $\delta_m(\vr)$ in terms of the $\vk$-space density field $\delta_{\vk}$, and write the expression for angular power spectrum, $\cl$, by using the definition of matter power spectrum $P(k)$, i.e. $\braket{ \delta_{\vk} \delta_{\vk'}} =(2\pi)^3 \delta^D(\vk -\vk') P(k)$ (for detailed formulation see \citealt{Loureiro:2019, Tiwari:2021DR1} etc.), 
\begin{eqnarray}
\label{eq:clth}
C_\ell&=&<|\al |^2> \nonumber\\
          &=&   \frac{2}{\pi }\int dk k^2 W^2(k) P(k,z=0)  \;,
\end{eqnarray}
 where $W(k)$ is the $k$-space window function given by
\begin{equation}
W^2(k)=  \left\vert \int_{0}^{\infty} d z D(z) b(z) N(z)  j_\ell(kr)\right\vert^2 
\end{equation}
where $D(z,k)$ represents the growth factor with dependence on $k$ due to deviations from the linear evolution on small scales.

We use the Core Cosmology Library (CCL\footnote{\href{https://github.com/LSSTDESC/CCL}{https://github.com/LSSTDESC/CCL}}, \citealt{Chisari:2019}) to calculate theoretical $\cl$.
To derive  $\cl$ from the masked data, we use the pseudo-$C_\ell$ recovery algorithm  by \cite{Alonso:2019}. The python modules for these algorithms are publicly available as {\tt pyccl}\footnote{\href{https://pypi.org/project/pyccl/}{https://pypi.org/project/pyccl/}} and  {\tt NaMaster}\footnote{\href{https://namaster.readthedocs.io/en/latest/index.html}{https://namaster.readthedocs.io/en/latest/index.html}}. Galaxies  are treated as discrete point sources and hence  the measured angular power spectrum of a galaxy catalog, $\cl^{\rm measured }$ ,  contains the  Poissonian shot-noise equal to $\frac{1}{\bar \cN}$. The $\cl^{\rm obs}= \cl^{\rm measured } - \frac{1}{\bar \cN}$ is equivalent to theoretical $\cl$ in Equation \ref{eq:clth}. The expected uncertainty in  $C_\ell$ determination due to cosmic variance, sky coverage, and shot-noise is, 
\begin{equation}
\label{eq:dcl}
\Delta C_\ell = \left( \frac{2}{(2\ell+1)f_{\rm sky}}\right)^{1/2} C^{\rm measured}_\ell,
\end{equation}
where $f_{\rm sky}$ represents the fraction of the sky covered by the survey. 

\subsection{Biasing scheme for AGNs}
\label{ssc:bias}
We assume that the AGNs catalog derived from CatWISE2020 follows a biasing scheme similar to radio-loud AGNs. This assumption may not hold entirely as for the WISE selected AGNs, we only have $\sim 42\%$ radio matches \citep{Stern:2012}. Nonetheless we use the same biasing scheme as for AGNs  from \cite{Adi:2015nb} and write 
\begin{equation}
\label{eq:bias}
b(z)=\frac{\int^{M_{\rm max}}_{M_{\rm min}} n_h(M,z) b_h(M,z) \mathcal{F}(M_*,z) dM }{\int^{M_{\rm max}}_{M_{\rm min}} n_h(M,z) \mathcal{F}(M_*,z) dM}  ,  
\end{equation}
where $n_h(M,z)$ is the halo number density for halos in  mass range $M$ and $M+dM$ at redshift $z$,  $b_h(M,z)$ is  the bias for these halos. $\mathcal{F}(M_*,z)$  represents the dependence 
of the fraction of infrared bright AGNs, $f_{IB}$, on the  stellar mass $M_*$. 
We assume  $f_{IB}$ to be of  the form
\begin{equation}
\label{eq:f_rl}
    f_{IB}\propto \mathcal{F}(M_*,z) = \left(\frac{M_*}{10^{11}M_{\odot}}\right)^{\alpha_0+\alpha_1 z}\; ,
\end{equation}
where $\alpha_0=2.5\pm 0.2$ obtained by \cite{Best:2005} with local ($z<0.1$) AGNs. The  parameter $\alpha_1$ contains the {infrared brightness} evolution as a function of redshift (or time). \cite{Adi:2015nb} find $\alpha_1=1.85\pm0.74$ using NVSS AGNs.  For our analysis, we use HMF: Halo Mass Function calculator \citep{Murray:2013} to obtain  $n_h(M,z)$ and $b_h(M,z)$, the python module of the code is publicly available as 
{\tt hmf}\footnote{\href{https://github.com/halomod/hmf}{https://github.com/halomod/hmf}}.

 The $M_*$ in equation \ref{eq:bias} is a function of halo mass  following the stellar-to-halo mass (SHM) relation from \cite{Moster:2013}. We choose $M_{\rm min}=4\times 10^{11} M_\odot$ corresponding to $M_*=10^{10} M_\odot$ which is the lowest stellar mass limit to host radio AGN \citep{Moster:2013}. Considering a lower value for $M_{\rm min}$ does not make a difference due to a very steep dependence of $\mathcal{F}$ on $M_*$. The upper limit for the integration, i.e., $M_{\rm max}$ is set to be $10^{15} M_\odot$, the halos above this mass are very rare, i.e., $n_h(M,z) \to 0$ for $M>10^{15} M_\odot$.

\subsection{Angular correlation function}
\label{sc:2pcf}
The angular correlation function, $w(\theta)$, is a measure of galaxy clustering in angular space;  an alternate to angular power spectrum discussed in Section \ref{ssc:cls}. The two-point angular correlation function is defined as the excess  probability, relative to  a random distribution,  of finding a pair of galaxies in both of the elements of solid angle $d\Omega_1$ and $d\Omega_2$ at angular separation $\theta$, 
\begin{equation}
\label{eq:wth}
dP= \bar \cN^2  (1+w(\theta)) d\Omega_1 d\Omega_2 ,  
\end{equation}
where $\bar \cN$ is the projected mean galaxy number density. To estimate w($\theta$) we employ \cite{Landy:1993} estimator which is  based on galaxy-galaxy (DD), random-random (RR), and galaxy-random (DR) pair counts and defined as, 
\begin{equation}
\label{eq:Landy-Szalay}
w(\theta)=\frac{DD-2DR+RR}{RR}
\end{equation}
In practice, we use the {\tt TreeCorr}\footnote{\href{http://github.com/rmjarvis/TreeCorr/}{http://github.com/rmjarvis/TreeCorr/}} python module by \cite{Jarvis:2004} to compute $w(\theta)$ from data. We derive the theoretical estimate of $w(\theta)$ from the model $\cl$ according to the relation 
\begin{equation}
\label{eq:cl2wth}
w(\theta)=\frac{1}{4 \pi}\sum_l (2l+1) \cl P_l(\cos \theta ), 
\end{equation}
where $P_l(x)$ are the Legendre polynomials. 

\section{Covariance matrix estimation and MCMC sampling }
\label{sc:mocks}
\begin{figure}
    \centering
    \includegraphics[width=0.55\textwidth]{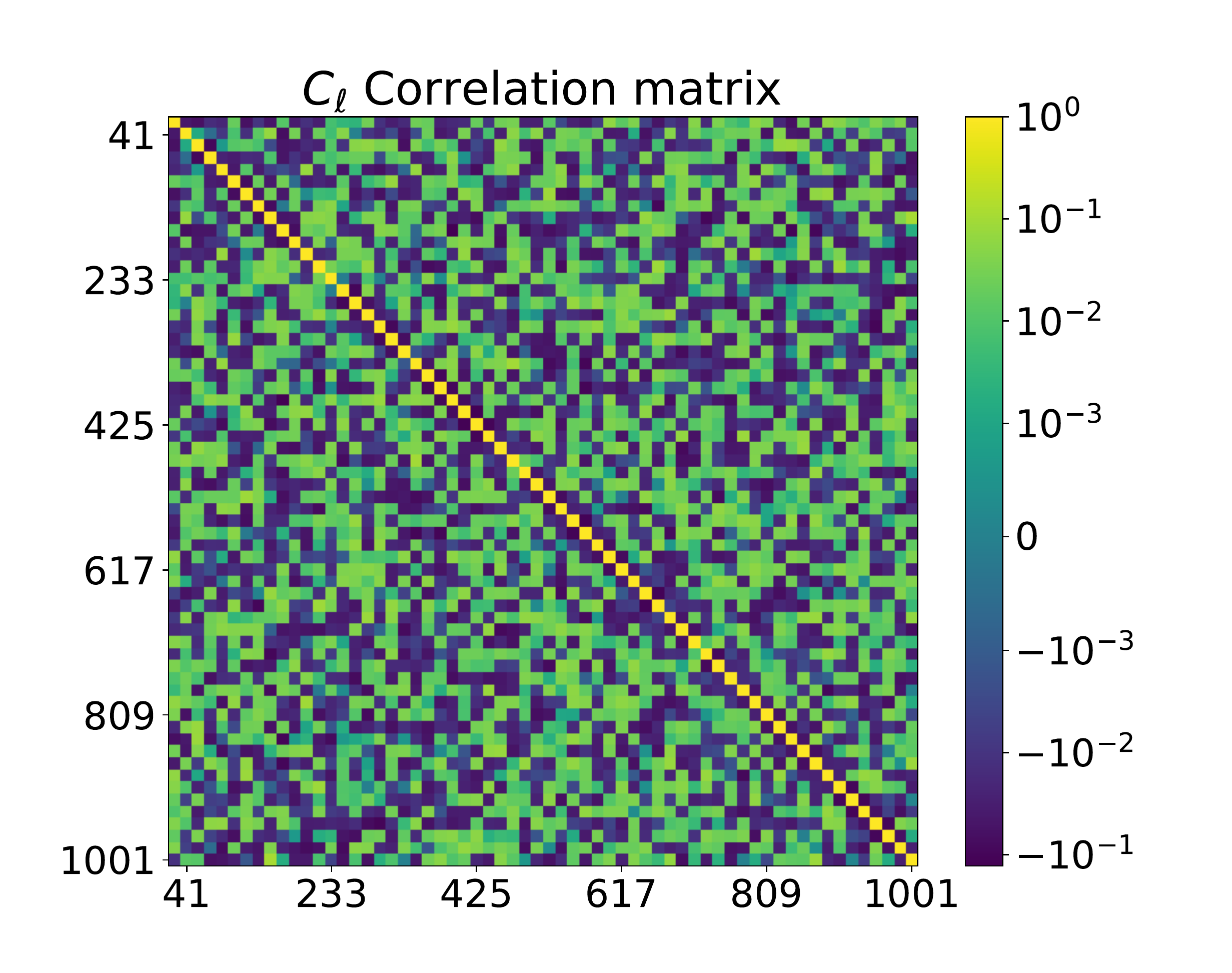}
    \caption{The angular power spectrum correlation matrix constructed from 1000 mocks generated using FLASK log-normal density simulator. Each galaxy mock follows the same sky coverage and contains approximately the same number of sources as our AGN-CatWISE2020 catalog.}
    \label{fig:covmat}
\end{figure}

For our pipeline calibration of data systematics and errors estimation, we generate $1000$  mocks of AGN-CatWISE2020 by employing the log-normal density field simulator code FLASK\footnote{\href{http://www.astro.iag.usp.br/~flask/}{http://www.astro.iag.usp.br/~flask/}} \citep{Xavier:2016}. We generate log-normal density fields  tomographically in $35$ redshift slices ($0$ to redshift $3.5$), each with a width $\Delta z=0.1$. We  neglect  AGNs beyond redshift 3.5 (see Figure \ref{fig:zpdf}). The statistical properties (i.e. auto and cross-correlations) of mocks are determined by the input angular power spectrum that we generate using {\tt CAMB} \citep{Challinor:2011}, including  the effect of  redshift-space distortions, non-linear power spectrum corrections, and lensing. For the mock generation, we assume the bias $b(z)$ and $N(z)$ distribution from \cite{Adi:2015nb} and generate auto and cross $C_\ell$s using {\tt CAMB} and provide inputs to FLASK to generate mock number count maps. We apply the survey mask,  shown in Figure \ref{fig:AGNcat}, to account for survey geometry. Next from the mocks, we calculate the angular power spectrum and compute the covariance matrix. To reduce noise and smooth recovery of pseudo-$\cl$ we collect power in multipole bins by averaging over 16 multipoles. We have shown the correlation matrix of the binned angular power spectrum in Figure \ref{fig:covmat}.

We fit the AGN-CatWISE2020  angular power spectrum  with standard $\Lambda$CDM and constrain the galaxy bias $b(z)$ and radial distribution $N(z)$. The likelihood,
 $\mathcal{L}(\clobs)$, i.e., the probability to observe data $\clobs$ given the model,   $ \mathcal{P}( \clobs|\bz,\nz) $ is 
\begin{equation}
   \mathcal{L} \propto \exp\left[-\frac{1}{2} (\vb{\clobs}-\vb{\clth})^{T} \vb{\Sigma^{-1}}(\vb{\clobs}-\vb{\clth})\right], 
\end{equation} 
where  $\vb{\Sigma}$ is the covariance matrix computed from the mocks discussed above. Applying  Bayes' probability theorem $P(A|B) P(B)= P(B|A) P(A)$, the model probability given the data, 
\begin{eqnarray}
\mathcal{P}( \bz,\nz|\clobs)=\mathcal{P}( \clobs|\bz,\nz) \times \nonumber \\ 
\mathcal{P}(\bz) \times \mathcal{P}(\nz)\; ,
\end{eqnarray}
  where $\mathcal{P}(\bz)$ and $\mathcal{P}(\nz)$ are the prior probability of bias $\bz$ and $\nz$, respectively. We consider parameterised $N(z) \propto z^{a_1} \exp\left[ - \left(\frac{z}{a_2} \right)^{a_3} \right]$ and the $\bz$ as discussed in Section \ref{ssc:bias}. The prior information for $\nz$ is the redshift histograms shown in Figure \ref{fig:zpdf}, and for $\bz$ we consider flat prior. For convenience we  use {\tt Cobaya}\citep{Torrado:2020xyz} to perform Bayesian analysis and {\tt CosmoMC} \citep{Lewis:2002ah,Lewis:2013hha} for MCMC sampling. 

\section{Results}
\label{sc:results}
\begin{figure}
    \centering
    \includegraphics[scale=0.6]{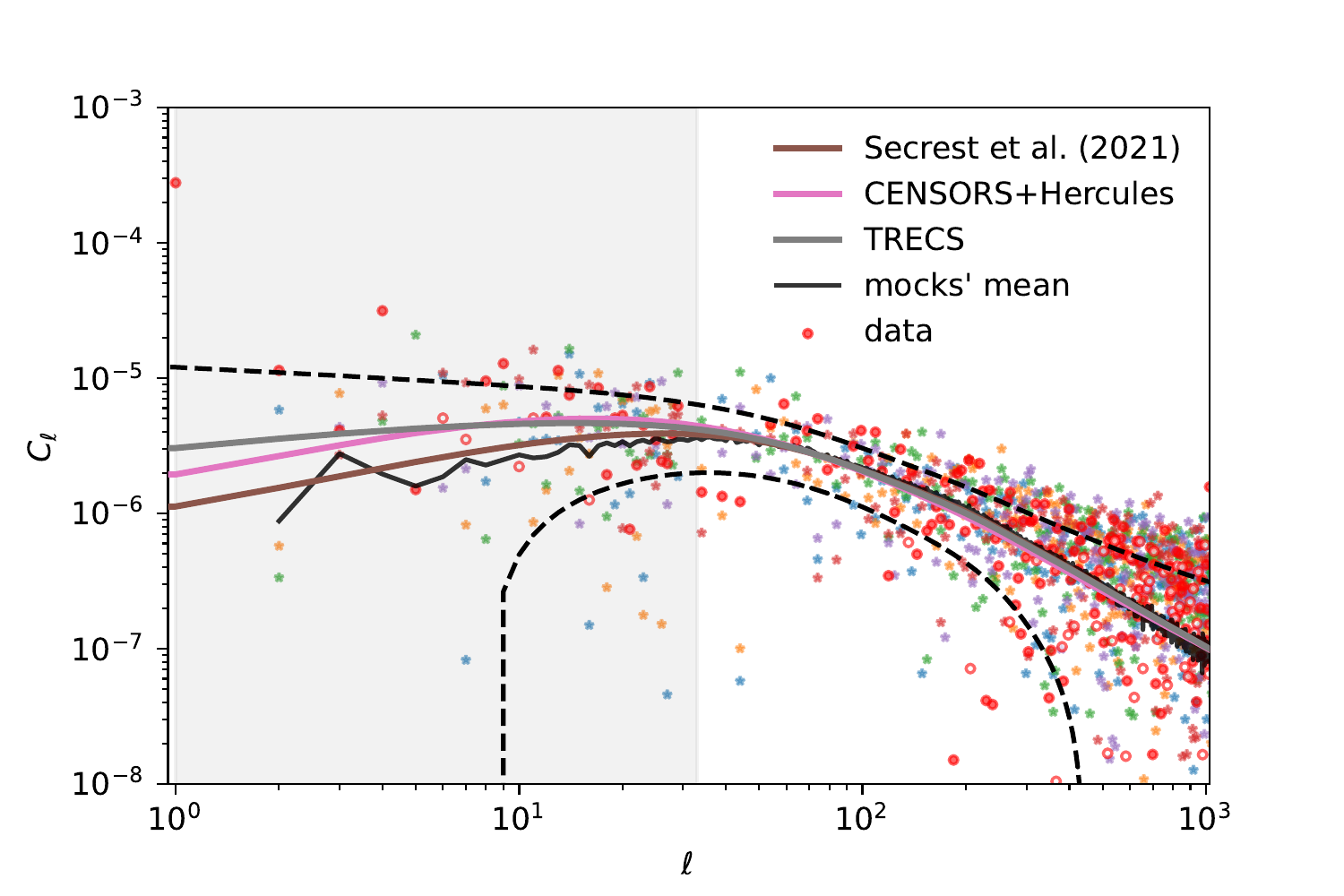}
    \caption{
    The angular power spectrum recovery from AGN-CatWISE2020. The open red circles are the data points with negative values (absolute value plotted). The dashed curves show the one sigma limits for the `TRECS' case, calculated using equation \ref{eq:dcl}.
    The data $\cl$s for $\ell \lessapprox 10$ are anomalous, showing excess power. We have reasonably good recovery above multipole $\ell \gtrapprox 30$, we only fit beyond the gray shaded area, and in this range, the mocks are a good match with the data. We have shown five random mocks (asterisks), and mocks' mean from all 1000 mocks for illustration. 
    The best fits with different $N(z)$ data are also shown.}
    \label{fig:raw_cls}
\end{figure}

\begin{figure}
    \centering
    \includegraphics[scale=0.6]{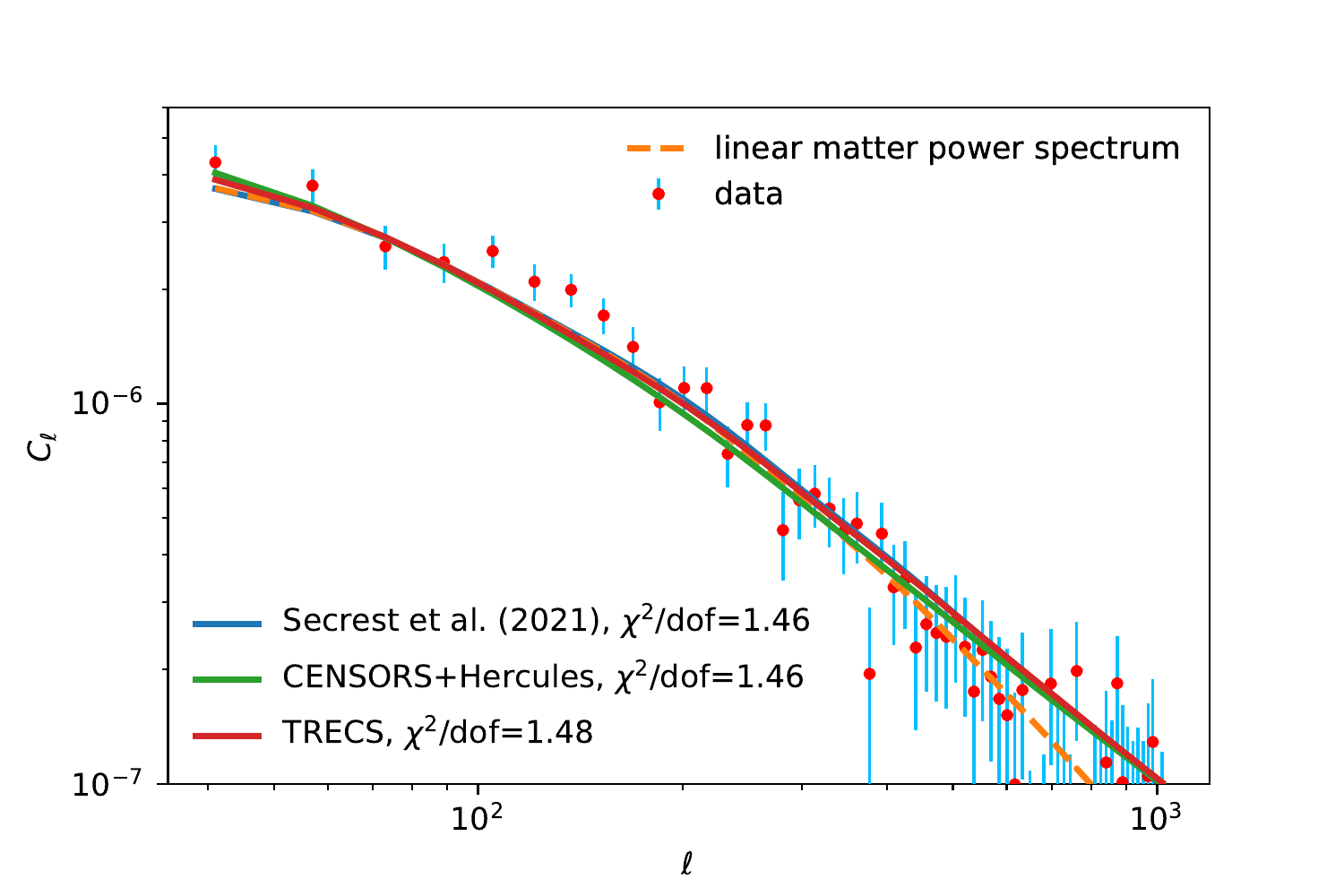}
    \caption{The angular power spectrum recovery in  multipole bands, averaging over groups of $\Delta \ell$ = 16 multipoles. The best fits with different $N(z)$ data are also shown. All fits are almost the same resulting $\chi^2/dof\approx 1.5$. The plotted error bars are the standard deviations from 1000 mocks.}
    \label{fig:cls}
\end{figure}

\begin{figure}
    \centering
    \includegraphics[scale=0.6]{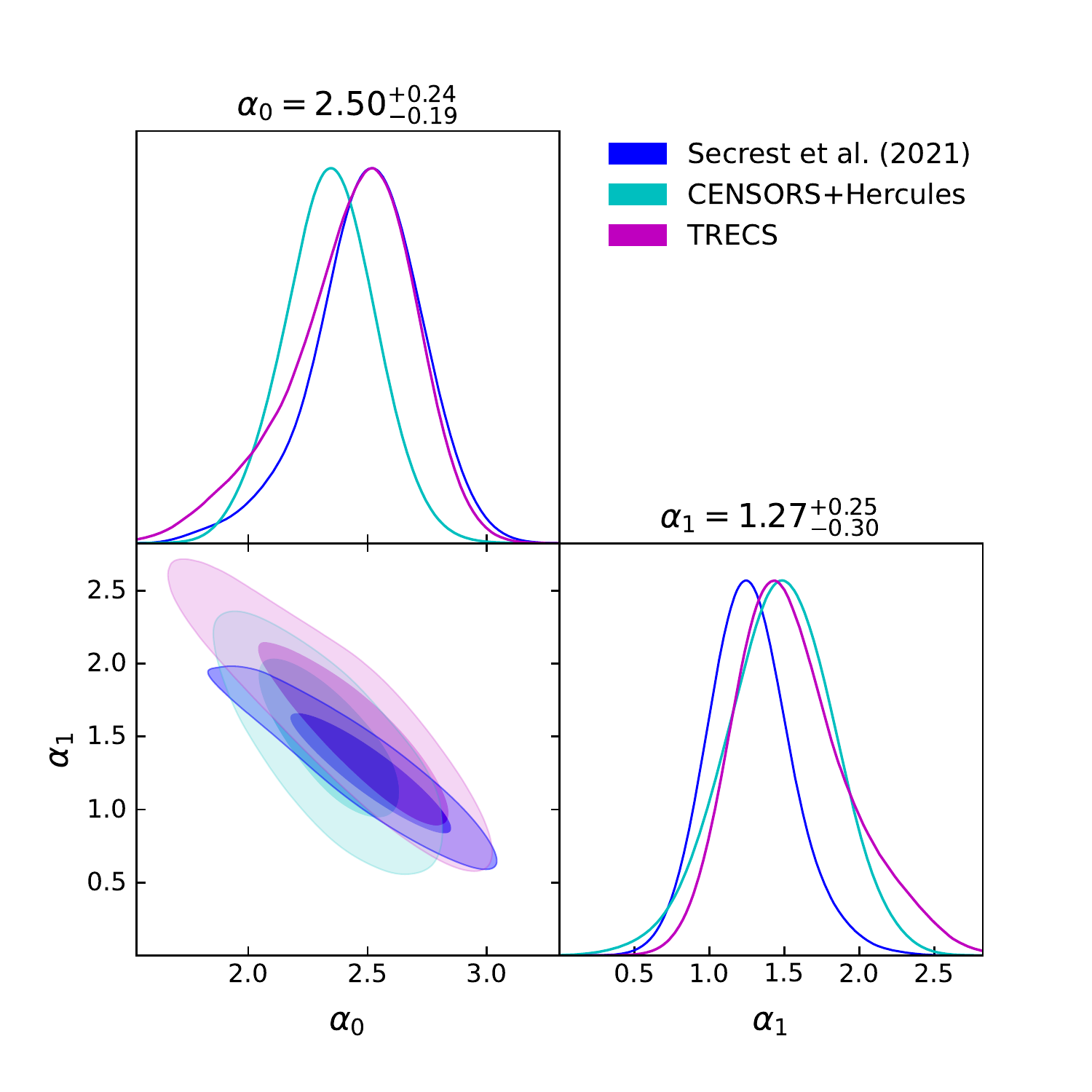}
    \caption{The best fitted $\alpha_0$ and $\alpha_1$  parameters assuming different $N(z)$ templates. The evolution parameter with $N(z)$ template from \cite{Secrest:2020CPQ} is  $\alpha_1=1.27^{+0.25}_{-0.30}$, this is $\approx 4.6 \sigma$ measurement of infrared bright AGN fraction evolution.}
    \label{fig:evolution}
\end{figure}

\begin{figure}
    \centering
    \includegraphics[scale=0.6]{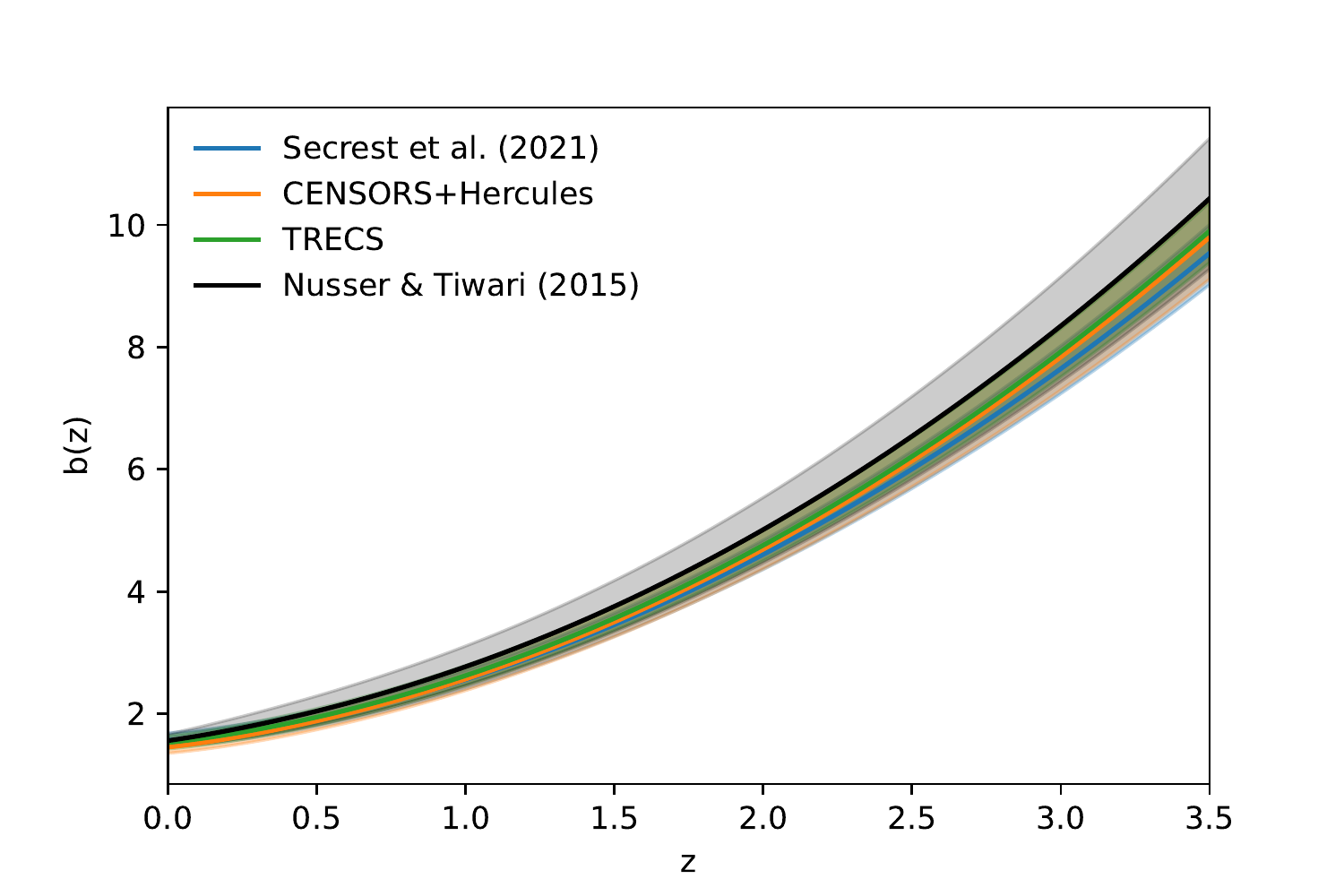}
    \caption{The best fit $b(z)$ and one $\sigma$ error bar.}
    \label{fig:bz}
\end{figure}

\begin{figure*}
    \centering
    \includegraphics[scale=0.6]{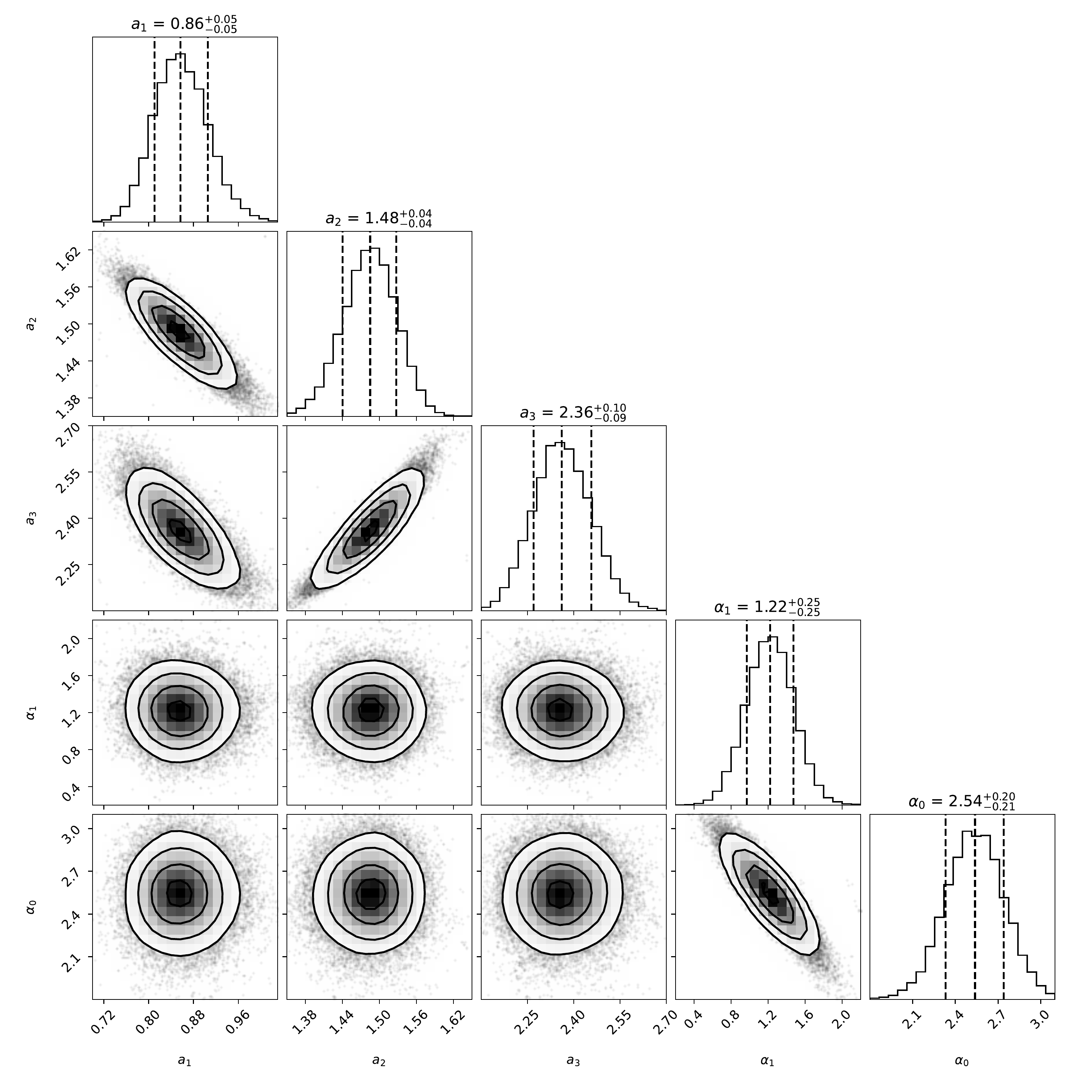}
    \caption{The MCMC samples and results assuming \cite{Secrest:2020CPQ} $N(z)$.  }
    \label{fig:cornerCatWise}
\end{figure*}

We recover the pseudo-$\cl$ from our masked AGN-CatWISE2020 (shown in Figure \ref{fig:AGNcat}) using {\tt NaMaster} \citep{Alonso:2019}. The results are shown in Figure \ref{fig:raw_cls}. We notice the excess power for multipoles,  $\ell \lessapprox 10$. The recovered $\cl$s beyond $\ell=30$ agree well  with the prediction of the standard $\Lambda$CDM cosmology. Also, the recovered pseudo-$\cl$s show a large scatter and therefore we use band-powers defined by averaging over bins of $\Delta \ell = 16$ multipoles. The results are presented in Figure \ref{fig:cls}.  
We consider the parameterized $N(z) \propto z^{a_1} \exp\left[ - \left(\frac{z}{a_2} \right)^{a_3} \right]$ and the bias $\bz$ as in equation \ref{eq:bias} and run MCMC sampling over parameters $a_1$, $a_2$, $a_3$, $\alpha_0$ and $\alpha_1$ to find maximum likelihood values of these for observed $\cl$s. We assume a flat prior for $\bz$ and for $\nz$ we consider the redshift distribution PDF by \cite{Secrest:2020CPQ}. We drop the largest scale $\cl$s and calculate the likelihood probability for the theory power spectrum in the multipoles range $\ell=33$ to $1025$. The non-linear $\Lambda$CDM power spectrum  fits the data well with $\chi^2/dof=1.46$. The non-linear fit is distinguishable from the linear matter power spectrum beyond $\ell \sim 250$\footnote{$\ell=250$ is around $0.7^\circ$ angular scale, assuming AGNs at redshift 1, this scale corresponds to $\approx 20$Mpc.}. The likelihood contours of the parameters $\alpha_0$ and $\alpha_1$ are shown in Figure \ref{fig:evolution} with best values  $\alpha_0=2.50^{+0.24}_{-0.19}$ and $\alpha_1=1.27^{+0.25}_{-0.30}$. These are consistent with \citep{Adi:2015nb} within one $\sigma$ error. The best $b(z)$ values are shown in Figure \ref{fig:bz}. A quadratic function $b(z)=1.54^{+0.13}_{-0.12}+0.53^{+0.01}_{-0.01} z+0.50^{+0.03}_{-0.03} z^2$ represents well the bias curve and its one sigma limits. The MCMC run samples, best values of the parameters, and parameter correlations are shown in Figure \ref{fig:cornerCatWise}. We add that the clustering dipole magnitude expected with this bias and $ N(z)$ assuming $\Lambda$CDM is $0.81\times 10^{-3}$ (i.e. $C_1=0.91\times 10^{-6}$)\footnote{Note the $C_1$ and dipole magnitude relation: $C_1 = \frac{4\pi}{9} |D|$ \citep{Gibelyou:2012}.}. However, notice that the dipole signal in galaxy count is largely from aberration and Doppler effects caused by local motion \citep{Ellis:1984}, and for CatWISE2020 the predicted value is $0.73\times 10^{-2}$ \citep{Secrest:2022}. 

We next employ redshift template produced using the TRECS simulation. We have created this template by matching the number density of AGN-CatWISE2020 and thus this arguably represents the redshifts PDF for the full sample. The TRECS redshifts fit the data well and we find $\chi^2/dof=1.48$. The fit to angular power spectrum is shown in Figure \ref{fig:cls}. We find slightly higher bias, $b(z)$, with TRECS (shown in Figure \ref{fig:bz}), even so the bias values remains consistent with \cite{Adi:2015nb}. 

We also consider redshift distribution from CENSORS and Hercules. The CENSORS survey is over 6 deg$^2$, and the Hercules is over 1.2 deg$^2$ both in total contain 165 sources above 7.5 mJy at 1.4 GHz. These surveys are presumably complete above 7.5 mJy at 1.4 GHz. The same redshifts were used to fit angular power spectrum from NVSS \citep{Adi:2015nb}. This $N(z)$ also fits the data equally well, $\chi^2/dof=1.46$, with bias  $b(z)$ slightly lower in comparison to the TRECS case.

\begin{figure}
    \centering
    \includegraphics[scale=0.6]{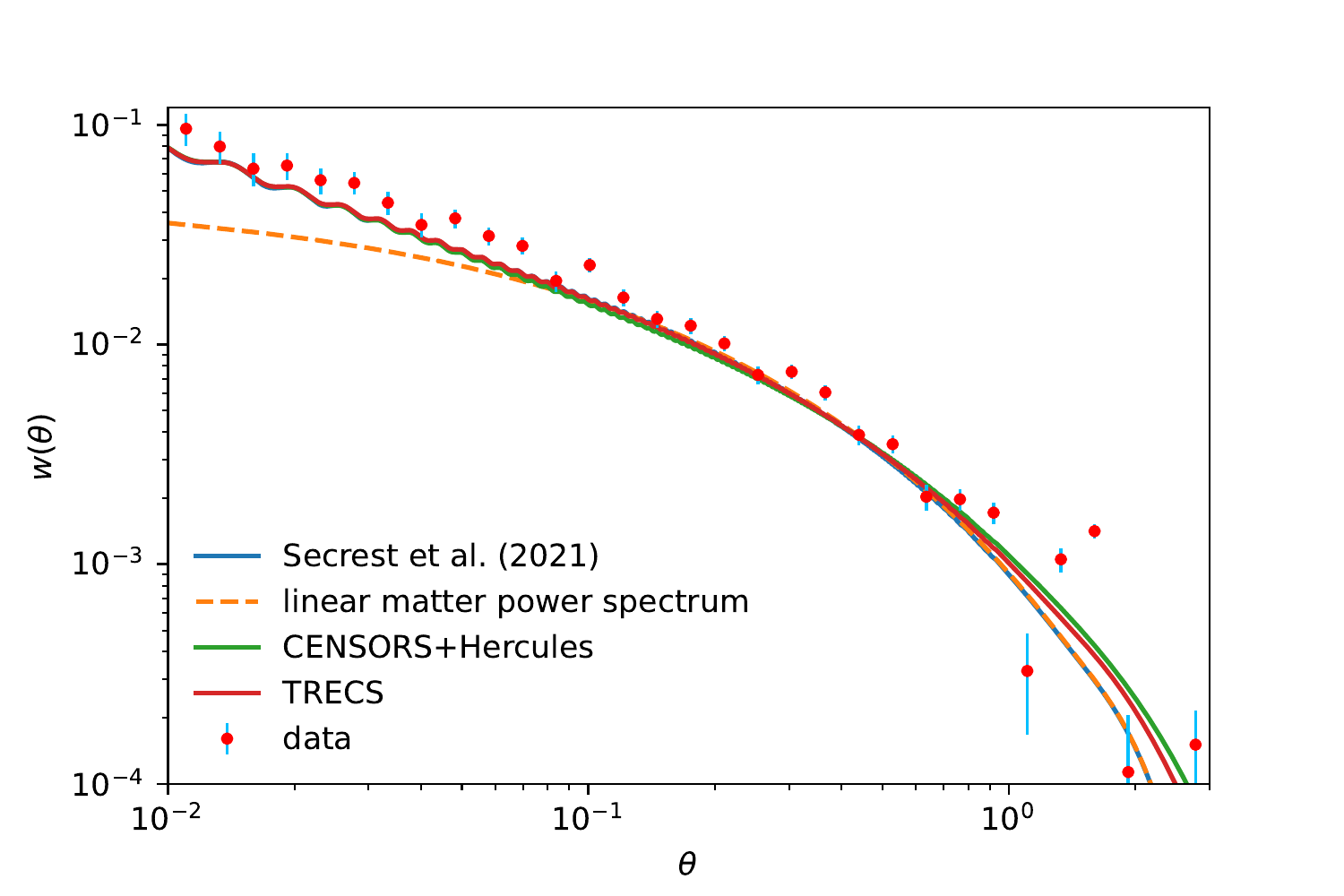}
    \caption{Angular two-point correlation function from AGN-CatWISE2020  and the theoretical fits. The fits are drawn using the best fit bias $b(z)$, and $N(z)$ obtained by fitting $\cl$s .}
    \label{fig:wtheta}
\end{figure}

The angular two-point correlation function represents the angular clustering measure in real space. We use the {\tt TreeCorr} \citep{Jarvis:2004} python module and calculate angular correlation function $w(\theta)$, the results are in Figure \ref{fig:wtheta}. We also plot the theoretical $w(\theta)$ obtained employing equation \ref{eq:cl2wth} and the best $\cl$ curves shown in Figure \ref{fig:cls}. The theoretical $w(\theta)$ plots show a reasonable match with data for scales corresponding to our multipoles fit range, i.e. $\ell<1025$ ($\theta \gtrapprox 0.18^\circ$). At scales below $0.1^\circ$ the best fit curves are moderately below the data points indicating slightly high clustering at very small scales.

\section{Discussion and Conclusion}
\label{sc:conclusion}
We have examined the angular clustering of the  AGNs in  CatWISE2020 catalog. 
The angular power spectrum  multipoles below $\ell \lessapprox 10$ are anomalous, with  $\cl$s that are systematically higher than the predictions of the $\Lambda$CDM. 
Nonetheless, the  $\cl$s  beyond $\ell\gtrapprox 30$ are in excellent agreement with the standard cosmology and we only use the $\cl$s  beyond $\ell>30$ to fit and obtain our results. Our analysis and results thus are independent of high dipole and quadrupole  and  any other systematics at the largest  scales. We reduce the measurements variance by binning the power spectrum, and average over groups of $\Delta \ell = 16$ multipoles, for a smooth recovery of angular $\cl$s. For error estimates we generated $1000$ mocks using FLASK \citep{Xavier:2016} and generate covariance matrix. 

We adopt  three redshift distribution templates and run MCMC to obtain the AGN clustering  bias factor as a function of redshift $\bz$. In particular, we use a $\nz$ prior information from \cite{Secrest:2020CPQ}, CENSORS+Hercules \citep{Rigby:2011, Adi:2015nb,Tiwari:2016adi}, and TRECS \citep{Bonaldi:2018}. The $\nz$ from  \cite{Secrest:2020CPQ} and  CENSORS+Hercules represent partial redshifts for AGN-CatWISE2020 sample and  could be closely representing  the redshifts distribution for the full sample too. Assuming the AGN-CatWISE2020 contains the radio bright AGNs, we run TRECS simulations to produce redshifts. This probably represents the redshifts for the full AGN-CatWISE2020 sample. We find that all three $\nz$ considered fit the data equally well producing $\chi^2/dof \approx 1.5$ with slightly different bias $\bz$. Therefore, the angular power spectrum fails to break  the degeneracy between $\bz$ and $\nz$, it effectively constrains only the combination of $\bz$ and $\nz$. 

We further explore the non-linear region of structure formation with AGN-CatWISE2020. The multipoles above $\ell \gtrapprox 250$ show a better match with the non-linear in comparison to the linear background matter power spectrum. With two-point correlation plot the difference between linear and non-linear power spectrum is clearly visible (Figure \ref{fig:wtheta}). At smallest scales the theoretical curves with non-linear power spectrum are clearly a better fit.

Remarkably, the AGN bias $\bz$ observed with NVSS (radio-selected) power spectrum in the linear regime of angular power spectrum $\ell\leq100$, agrees within its one sigma limit with AGNs in CatWISE2020 infrared-selected sample, for liner and non-linear ($\ell\gtrapprox250)$ regimes. Although, we obtain slightly smaller bias for infrared-selected AGNs. This consistent with the results obtained by \cite{Hickox:2009ApJ}. However, note that the extracted bias value depends on $N (z)$ being assumed, and we expect to be more certain about bias values once we have better redshift measurements.

We also explore  the infrared brightness dependence of AGNs on star formation by considering the the radio luminosity dependence on stellar mass. Adopting the form  $f_{\rm IB} \sim M_*^{\alpha_0+\alpha_1 z}$ for the fraction of infrared bright galaxies versus stellar mass, we measure the evolution parameter, $\alpha_1$. Assuming a prior $\nz$ from \cite{Secrest:2020CPQ}, we derive $\alpha_1=1.27^{+0.25}_{-0.30}$. Slightly larger values, $\alpha_1=1.45\pm0.38$ and $1.48\pm0.28$ are obtained with  $\nz$ priors from CENSORS+Hercules and TRECS, respectively. It is interesting to note that the biasing scheme and $\nz$ of radio-selected  AGNs work for infrared AGNs clustering signal. There may be some alternatives to biasing scheme formulation, nevertheless, the resultant $\bz$ shown in Figure \ref{fig:bz} is robust and immune to biasing schemes. A quadratic bias $b(z)=1.54+0.53z+0.50 z^2$ conveniently represent the best bias for CatWISE2020 AGNs.

We also calculate angular two point correlation function from data and obtain reasonable recovery up to a few degrees. The theoretical curves obtained from $\cl$s,  are a good match to data. The angular distribution of AGNs from CatWISE2020 above multipole $\ell\gtrapprox 10$ is smooth and a reasonable fit with standard cosmology.

\section{Acknowledgments}
 We thank Subir Sarkar and Sebastian von Hausegger for providing the redshifts distribution prepared in  \cite{Secrest:2020CPQ}. PT acknowledges the support of the RFIS grant (No. 12150410322) by the National Natural Science Foundation of China (NSFC). GBZ is supported by the National Key Basic Research and Development Program of China (No. 2018YFA0404503), NSFC Grants 11925303 and 11720101004. This research is supported by a grant from the Israeli Science Foundation grant number 936/18 and the Technion Asher Space Research Institute.
 \software{HEALPix \citep{Gorski:2005}, Core Cosmology Library \citep{Chisari:2019}, NaMaster \citep{Alonso:2019}, hmf \citep{Murray:2013}, TreeCorr \citep{Jarvis:2004}, FLASK \citep{Xavier:2016}, Cobaya \citep{Torrado:2020xyz}, CosmoMC \citep{Lewis:2002ah,Lewis:2013hha}, CAMB \citep{Challinor:2011}.}

\bibliographystyle{aasjournal}
\bibliography{main}
\end{document}